\documentclass{vgtc}                          

\ifpdf
  \pdfoutput=1\relax                   
  \usepackage{graphicx}                
  \DeclareGraphicsExtensions{.pdf,.png,.jpg,.jpeg} 
\else
  \usepackage{graphicx}                
  \DeclareGraphicsExtensions{.eps}     
\fi%

\graphicspath{{figures/}{pictures/}{images/}{./}} 

\usepackage{microtype}                 
\PassOptionsToPackage{warn}{textcomp}  
\usepackage{textcomp}                  
\usepackage{mathptmx}                  
\usepackage{times}                     
\usepackage{cite}                      
\usepackage{tabu}                      
\usepackage{booktabs}                  

\usepackage{subfig}
\usepackage{csquotes}
\usepackage{algorithm}
\usepackage{algpseudocode}
\usepackage{dblfloatfix}
\usepackage{url}
\usepackage[]{hyperref}

\makeatletter
\newcommand{\thickhline}{%
	\noalign {\ifnum 0=`}\fi \hrule height 1pt
	\futurelet \reserved@a \@xhline
}
\newcolumntype{"}{@{\hskip\tabcolsep\vrule width 1pt\hskip\tabcolsep}}
\makeatother

\onlineid{1272}

\vgtccategory{Research}

\vgtcinsertpkg

\title{Assessment HTN (A-HTN) for Automated Task Performance Assessment in 3D Serious Games}

\author{
	Kevin Desai\thanks{email: kevin.desai@utsa.edu}\\
    \scriptsize Department of Computer Science\\
    \scriptsize The University of Texas at San Antonio
    \and Omeed Ashtiani\thanks{email: omeed.ashtiani@utdallas.edu}\\
    \scriptsize Department of Computer Science\\
    \scriptsize The University of Texas at Dallas
    \and Balakrishnan Prabhakaran\thanks{email: bprabhakaran@utdallas.edu}\\
	\scriptsize Department of Computer Science\\
	\scriptsize The University of Texas at Dallas
}

\abstract{
In the recent years, various 3D mixed reality serious games have been developed for different applications such as physical training, rehabilitation, and education.
Task performance in a serious game is a measurement of how efficiently and accurately users accomplish the game's objectives.
Prior research includes a graph-based representation of tasks, e.g. Hierarchical Task Network (HTN), which only models a game's tasks but does not perform assessment.
In this paper, we propose Assessment HTN (A-HTN), which both models the task efficiently and incorporates assessment logic for game objectives.
Based on how the task performance is evaluated, A-HTN automatically performs: (a) Task-level Assessment by comparing object manipulations and (b) Action-level Assessment by comparing motion trajectories.
The system can also categorize the task performance assessment into single user or multi-user based on who is being assessed. 
We showcase the effectiveness of the A-HTN using two 3D VR serious games: a hydrometer experiment and a multi-user chemistry experiment.
The A-HTN experiments show a high correlation between instructor scores and the system generated scores indicating that the proposed A-HTN generalizes automatic assessment at par with Subject Matter Experts.
} 

\CCScatlist{
  \CCScatTwelve{Human-centered computing}{Mixed / augmented reality}{};
  \CCScatTwelve{Human-centered computing}{Collaborative interaction}{};
  \CCScatTwelve{Applied computing}{Interactive learning environments}{};
  \CCScatTwelve{Social and professional topics}{Student assessment}{};
  \CCScatTwelve{Human-centered computing}{User models}{}
}

\begin{document}



\maketitle

\section{INTRODUCTION} \label{introduction}                                              
3D games have started being utilized for virtual laboratories with the purpose of educating and imparting practical knowledge in the science, technology, engineering, and mathematics (STEM) fields. 
Due to the recent COVID pandemic, many universities have begun offering virtual, at home courses rather than in person courses. Virtual reality classrooms have been embraced by institutions such as the University of New Hampshire i.e. their Parpak Project. \cite{parpakNewHampshire}. 
Many of these classroom environments, such as a chemistry laboratory course, require 3D space to fully manipulate necessary tools. 
For example, consider a chemistry hydrometer experiment in which the student needs to place the fragile hydrometer in a cylinder by making sure that the hydrometer does not touch the walls of the cylinder. 
In such a case, perception of depth is vital, requiring that a simulation of this experiment is designed as a 3D environment.
As the adoption of such 3D virtual environments increase, the need for automatic assessment of the students' competency within that increases as well.

\begin{figure}[]
	\centering
	\subfloat[]{ \includegraphics[width=80mm, keepaspectratio=true]{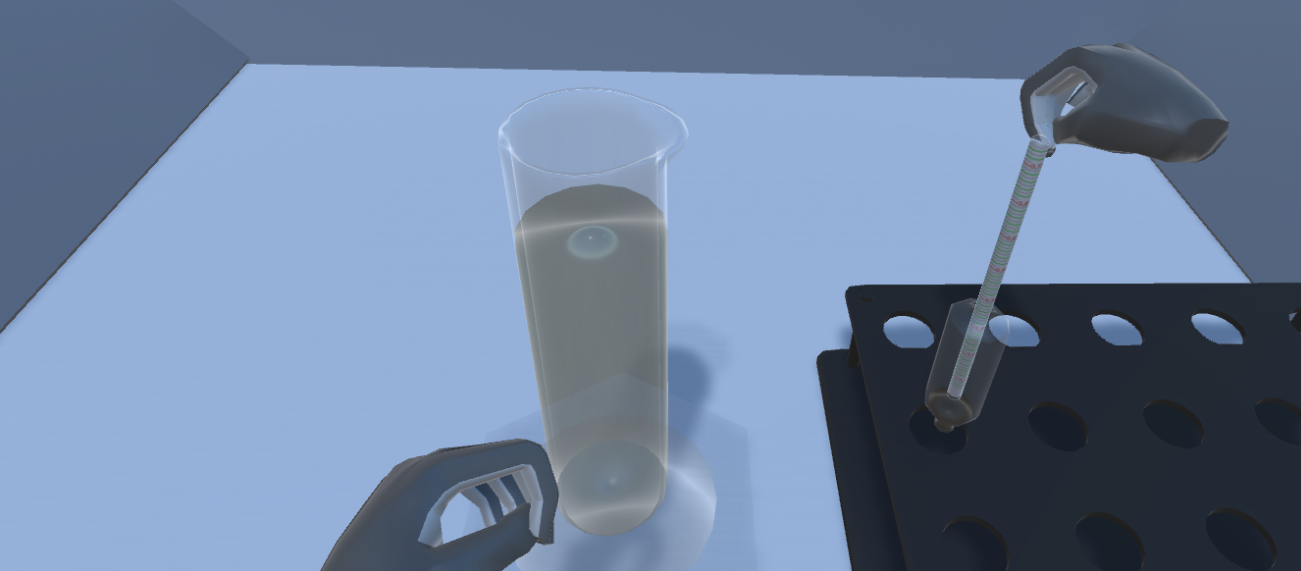} \label{fig:hydroExp}} \\
	\vspace{-2mm}
	\subfloat[]{ \includegraphics[width=80mm, keepaspectratio=true]{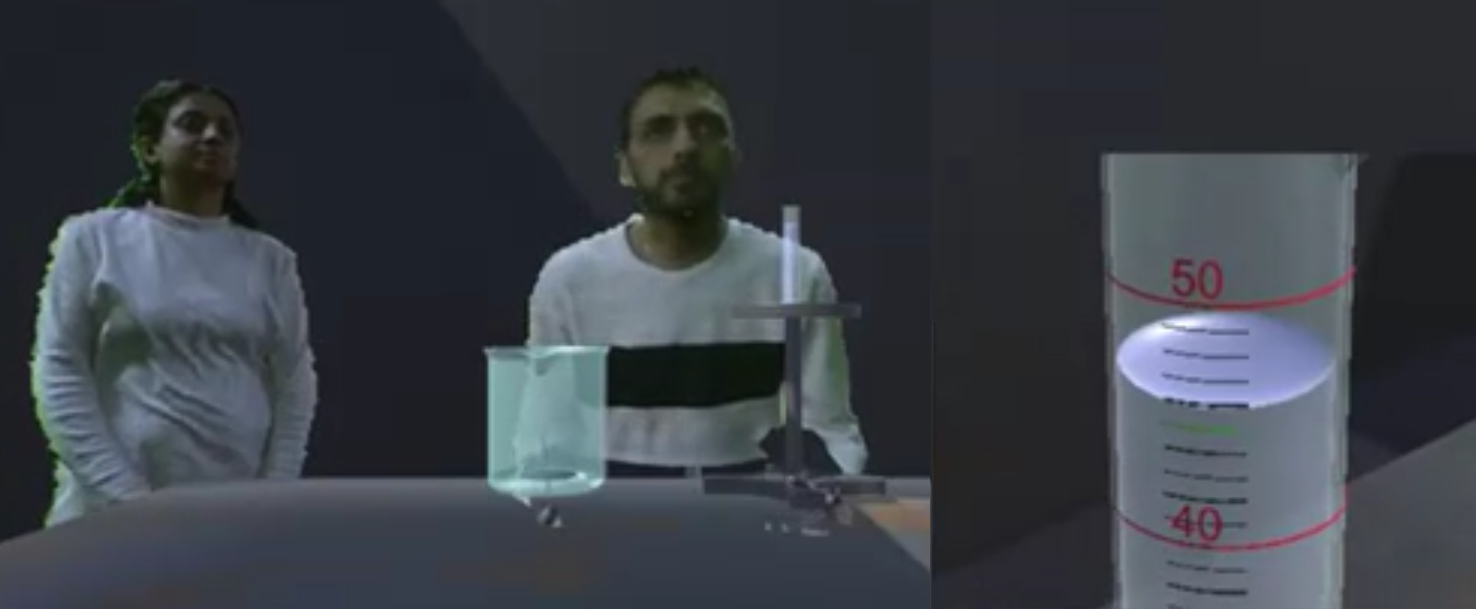} \label{fig:collabExp}} 	\caption{Serious Game Case Studies - (a) Hydrometer Experiment (b) Collaborative Chemistry Experiment}
	\vspace{-2mm}
	\label{fig:caseStudies}
\end{figure}

Task Performance in a serious game refers to how effectively and accurately a user performs tasks to achieve an overall objective.
Assessment of a user's task performance in a game requires incorporation of the fine-grained details of each task in the game design. 
In the real world, a Subject Matter Expert (SME), e.g., therapist or instructor, is usually present to assess the user's task performance.
In order to create automatic assessment, the game author should work with an SME and use a principled assessment design framework while incorporating the fine-grained details of each task. 
One such framework, known as Stealth Assessment \cite{saCompLearn}, measures the conceptual construct for competency (i.e., skills and abilities) and in-game tasks as well as the evidence of the competency captured during game play.

\subsection{Task Performance Assessment Categorization} \label{assessCat}

Serious games, such as those used for STEM training, may be single player or multi-player. Depending on \textit{`who'} is being assessed, the task performance assessment can be categorized into:
\begin{itemize}
	\item \textbf{Single user} - assesses the serious game task performance of a user in a single player environment.
	\item \textbf{Group} - assesses multiple people in a serious game as a whole; assesses an entire group at a time.
	\item \textbf{Individual user in the group}- assesses the serious game task performance of a single user in a multi-player environment.
\end{itemize}

In addition to \textit{`who'} is being assessed, we also need to know \textit{`how'} the assessment is to be performed.
In robotics, Learning from Demonstration (LfD) systems have been designed for non-experts to teach a robot to carry out a new task.
As explained in \cite{LfD}, LfD approaches can typically be considered under two broad categories: (a) \textit{Task-level learning:}\cite{taskLearning2, taskLearning3, taskLearning5} high-level complex tasks such as stacking blocks or sorting objects, assuming robots have a knowledge of the basic motions that goes with these tasks, and (b) \textit{Action-level learning:}\cite{actionLearning2, actionLearning3, actionLearning4} teaching robots low-level manipulation trajectories such as grasping or picking-up objects.
We categorize task performance assessment into:
\begin{itemize}
	\item \textbf{Task-level Assessment:} User's task performance is assessed by comparing the different virtual \textbf{object manipulations} - position, orientation, collision, attachment, text input, etc. This is vital for evaluating the competency of the user and knowledge of the tasks.
	\item \textbf{Action-level Assessment:} Instead of comparing just the start or final positions, the \textbf{motion trajectory} of the objects is used to assess the task performance. This is useful for understanding certain safety practices, i.e., holding test tubes properly, the distance of a chemical from the face, or passing a tool to another user.
\end{itemize}

\subsection{Proposed Approach} \label{propApp}
To perform automated task performance assessment in serious games, we need to represent both the task and the assessment logic.
Initial use of modeling strategies for serious games have been proposed using Hierarchical Task Network (HTN) \cite{htnTireChange} and Dynamic Bayesian Network (DBN) \cite{dbnAssess}.
However, the scope of the automated assessment is very limited, mainly to a specific serious game performed by a single user \cite{petriNetAirTraffic} or in online 2D games \cite{searchAlgo2dEval, sietteAssess, moocAssess}.
To the best of our knowledge, multi-user automated task performance assessment in 3D serious games has not been explored.
The HTN model \cite{htnAnsi} is only used for graphical representation of tasks and not performance assessment.
In this paper, we propose the Assessment HTN (A-HTN) which represents not only the tasks but also carries out automated task performance assessment.
With this augmentation, we also expand the scope of the serious game tasks and performance assessment to any number of users, not just one.
A-HTN should work with any type of game, such as 1D, 2D, or 3D, yet we showcase the effectiveness of A-HTN by using 3D virtual reality-based serious game case studies, as shown in Figure \ref{fig:caseStudies}.

\section{RELATED WORK} \label{RW}
Significant research has been performed for modeling the tasks in a serious game.
\cite{GIFT-survey} provides a background on the different uses of intelligent tutoring systems (ITSs) in context of course instruction and educational research. 
Generalized Intelligent Framework for Tutoring (GIFT) is provided as an open-source, domain independent ITS framework for creating adaptive tutoring content for classroom use.
Authoring tools for instructors and Subject Matter Experts (SMEs), which do not require a background in computer science to use, are provided by GIFT to create fully adaptive computer-based lessons.
The components of the lesson, such as surveys, quizzes, lesson materials and videos, as well as the path of the lesson can be determined by the instructors.
\cite{ITSsurvey} performed a systematic literature review of 33 papers, which were finalized from a list of 4622 papers retrieved from seven digital libraries published from 2009 to 2013, to understand how authoring tools have been used by non-programmer authors for ITS design.
\cite{sietteAssess} explained the domain-independent webbased Siette assessment environment that can be semantically integrated with intelligent systems or with large Learning Management Systems (LMSs), such as Moodle. 
It incorporated different scaffolding features, such as hints, feedback, and misconceptions as well as other features covering different educational needs and techniques, such as spaced repetition, collaborative testing, or pervasive learning.
For massive online open courses (MOOCs), \cite{moocAssess} proposed a learner cold-start algorithm for sampling a group of initial, diverse questions, based on determinantal point processes, to provide useful feedback to the learner. 
\cite{crowdAssessRubrics} mentioned an automated approach for developing reliable rubrics for educational intervention studies that address reading and writing skills.
It compared a main idea rubric used in a successful writing intervention study to a highly reliable wise-crowd content assessment method developed to evaluate machine-generated summaries.
An educational system with integrated assessment approaches was used by tutors to assess performance by 400 students in learning search algorithms \cite{searchAlgo2dEval}.

\begin{figure*}[]
	\centering
	\includegraphics[width=160mm, keepaspectratio=true]{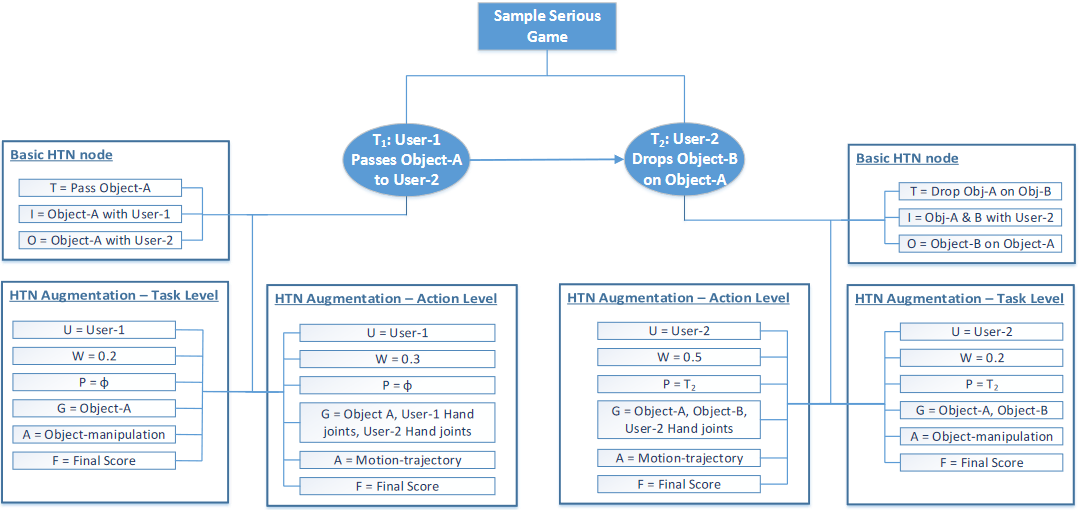}
	\caption{Assessment HTN (A-HTN) for the sample serious game}
	\label{fig:AssessmentHTN}
\end{figure*}


To our knowledge, only a few modeling and learning strategies have been explored to formally represent 3D serious games, such as Chemistry laboratories \cite{kevinMMCVL}, and to incorporate assessment. 
Although the following models were not used in 3D virtual reality games, they do incorporate assessment. 
Hierarchical Task Network (HTN) is a graphical representation for tasks performed in a game, with formalization as an ANSI/CEA-2018 standard \cite{htnAnsi}.
\cite{htnTireChange} used HTN for interactive hierarchical task learning from a single demonstration for an example application of changing tires.
\cite{petriNetAirTraffic} showed a colored PetriNet based model that is used to assess a human operator’s cognitive performance in an air traffic control scenario. 
\cite{dbnAssess} and \cite{dynamicBaysNets} use hierarchical Dynamic Bayesian Networks to introduce a computational framework incorporating automatic performance assessment of complex tasks or action sequences. 
\cite{dynamicBaysNets} uses hierarchical Dynamic Bayesian Networks to model student knowledge regarding five different intelligent tutoring systems and perform stealth assessment to adapt the needs of individual students.
\cite{studentCompetency} goes further and compares different models such as single-task and multi-task models, both static and sequential using the stealth assessment framework. 
\cite{deepStealth} also uses the stealth assessment framework with deep neural networks that are pre-trained using stacked denoising autoencoders, outperforming classical machine learning techniques such as support vector machines and naïve Bayes.
Similarly, \cite{lstmStealth} utilizes long-short term memory networks to improve stealth assessment. 
Stealth assessment is grounded in Evidence Centered Design (ECD), which features task, evidence, and competency models to conduct probabilistic reasoning about knowledge, skills, and abilities of students \cite{focusArticle}. 
In this work, we utilize the ECD framework with a Hierachical Task Network (HTN) to create the Assessment HTN (A-HTN) for automated task performance assessment.

\textbf{Evidence Centered Design (ECD)} is an assessment design framework that includes several elements to model the assessment process in a succinct and dependable manner. 
These elements are (1) the competency model: what knowledge and skills should be assessed, (2) the evidence model: what behaviors or performances should reveal these competencies, and (3) the task model: what tasks should elicit the behaviors \cite{saCompLearn}. 
The evidence model is the link between the competency and the task model, which observes the in-game tasks and the underlying competency constructs.
\cite{focusArticle} describes the general methodology for developing assessments, which we use in conjunction with Hierarchical Task Networks to create the Assessment HTN (A-HTN).
As we will see with the A-HTN, the competency model defines which knowledge and skills should be assessed, i.e. do the users understand the hydrometer experiment and do they understand safety precautions.
In order to display this competency, we measure the evidence of how closely the user follows the object manipulations and motion trajectory of the SME. 
The task model in the ECD model is directly correlated to our task-level and action-level assessment.

\section{ASSESSMENT HTN (A-HTN)} \label{systemDesign}
A Hierarchical Task Network (HTN) \cite{htnAnsi} can graphically represent tasks in games by using two types of nodes - Abstract and Primitive.
Abstract nodes are high level representations of the different tasks involved consisting of one or more primitive nodes and zero or more abstract nodes.
Primitive nodes are leaf nodes which represent a specific task to be performed in the game. 
The order of execution of the tasks is determined using arrows between the nodes.
Each primitive HTN node can represent three things: (i) $T=$ the task to be performed, (ii) $I=$ the input to the task and (iii) $O=$ the output of the task. 
Hence, we can represent an HTN primitive node as $N_i = \{T_i, I_i, O_i\}$.

Consider a game consisting of two users - $U_1$ and $U_2$, who collaborate and work on two objects - $A$ and $B$. 
There are two tasks to perform: $T_1$ where $U_1$ passes object-$A$ to $U_2$ and $T_2$ where $U_2$ drops object-$B$ onto object-$A$.
Task $T_2$ can only be performed once $T_1$ is completed. 
Task performance for each of the collaborative users, $U_1$ and $U_2$, needs to be assessed individually based on the virtual objects manipulated in each task.
The standard HTN representation cannot represent more than one user and cannot perform automated task performance assessment.

\subsection{Augmentation to HTN} \label{augHTN}
In order to incorporate task performance assessment for a multi-user serious game, we propose the Assessment HTN (A-HTN) in which the primitive node $N_i$ is augmented with the following parameters:

\begin{itemize}
	\item $U_i=$ The \textbf{User(s)} whose task performance will be assessed for this task. This will be one of the three assessment categories - single user, group, or individual user in a group.
	\item $W_i=$ The \textbf{Weight} of a specific task in the overall performance assessment for the selected user(s).
	\item $P_i=$ \textbf{Predecessor tasks}  that require completion before the start of this task. This is used to generate the proper hierarchical ordering of the tasks to be performed.
	\item $G_i=$ The \textbf{Game objects} that are modified and used to perform the assessment of this task.
	\item $A_i=$ The \textbf{Assessment} the type of assessment performed on this task. There are two types of assessment: task-level assessment using object manipulation, action-level assessment using motion trajectories, or a combination of both. The user's task performance is compared with the SME reference performance to assess the task and provide feedback.
	\begin{itemize}
		\item \textbf{Time} can be added as an extra performance feature for the comparison. Many serious games inherently have a timing constraint incorporated into the design. If the timing constraint is not inherent, it can be incorporated into the assessment parameter to perform a time-based assessment.
	\end{itemize}
	\item \textbf{$F_i=$ Feedback} provided to the user: either \textit{real-time} or \textit{final score}. \textit{Real-time} denotes assessing the user's performance and providing feedback in real time while the user is playing the game. \textit{Final-score} denotes assessing the user throughout the game and calculating a final score after task completion.
\end{itemize}

After augmenting the above parameters, a primitive A-HTN node $N_i$ can now be represented as $N_i$ = $\{T_i, I_i, O_i, U_i, W_i, P_i, G_i, A_i, F_i\}$.
To generate a complete A-HTN for a serious game, an SME is required to provide details for each task.
For each task, the SME first provides a \textit{task name} and selects the \textit{task type} as either abstract or primitive.
\textit{Abstract tasks} have sub-tasks that must be defined and hence will not have any other A-HTN parameter assigned. 
On the other hand, \textit{primitive tasks} require that all A-HTN parameters are assigned.
The SME also selects the \textit{predecessor tasks} $P_i$, required to be performed before the execution of the given task. 
This information is used to create the arrows between the A-HTN nodes. 
The assessment $A_i$ can be performed using \textit{task-level assessment}, \textit{action-level assessment}, or a combination of both.
The remaining A-HTN parameters are assigned differently with respect to the selected assessment strategy.

Figure \ref{fig:AssessmentHTN} shows the A-HTN for a sample serious game, including primitive nodes with augmented parameters.
The augmented parameters have both task-level and action-level assessment which are kept separate for clarity, allowing instructors or SMEs to check performance assessments independently should they wish to.

\subsection{Task-level Assessment} \label{taskLevelAssess}

Task-level assessment depends on which objects are manipulated and how.
To assess the user's performance, reference data is provided by the SME in the form of task performance recordings.
If only one reference recording is provided, it is considered the best way to perform the task with the worst case either being lack of movement or no change. 
The task of following the SME's trajectory follows as evidence for competency in understanding the experiment.
A quality rating, ranging from $0.0-1.0$, is also provided for each reference performance. 
For each task $T_i$ performed by the user in the serious game, a score value of $\Omega_i$ is calculated from $0.0 - 1.0$, depending on the relative closeness with the reference SME recording. 
Weights $W_i$ are provided by the SME and are used to compute the final score $\Delta$.

The proposed A-HTN can incorporate any type of object manipulation check if defined correctly with the help of an SME. 
The following checks we consider in our case studies are:
\begin{itemize}
	\item \textbf{Orientation} - a check that computes a score by comparing the average orientation of the selected object $G_j$ for the user's performance against the reference orientations provided by the SME.
	\item \textbf{Position}  - similar to the orientation check, compares the user's average position of object $G_j$ with the reference positions and computes a distance-based score.
	\item \textbf{Attachment} - ensures that object $X$ is always attached to another object $Y$. The score value is computed as the time duration ratio in which $X$ is attached to $Y$ with respect to the total time taken to perform the task.
	\item \textbf{Collision}  - ensures that object $X$ does not collide with another object $Y$ or another object present in the scene. Depending on the importance, a set penalty value, e.g. $0.01$, can be incurred on the score each time a collision occurs.
	\item \textbf{Text value input} - computes a score by comparing the user input value obtained using a keyboard against the reference values.
\end{itemize}

\subsection{Action-level Assessment} \label{actionLevelAssess}
For some types of serious games, e.g., the hydrometer experiment, motion trajectory is of significant important and thus we include an action-level assessment strategy. 
This importance comes from the proper handling and safety requirements as vials and tools should be handled with care, following proper protocol and safety precautions.
Action-level assessment focuses on the motion trajectory of the player rather than the task-level object manipulations. 
For each primitive task, the SME can provide reference to not only each item, but also the motion trajectory information recorded as skeleton and mesh files. 
The SME selects the $G$ body parts for the task that will be assessed, i.e. head, hand, and fingers, for which the user's skeletal positions are used to calculate user performance from reference data. 
We take the differences in body configuration of the user and SME into account by performing a correlation based correction using the height ratio between face and hands. 
During live user performance, visual feedback is provided by generating bubbles at the reference joint locations.
If the user can match the SME reference points for all of the $G_j$ skeletal points, the bubbles burst and the game transitions to the next frame.
If more than $skip-time$ has elapsed, e.g. $5$ seconds, we assume that the user is not able to make the appropriate movement and move on to the next movement by updating the bubble positions, if applicable.
Live feedback is also provided in terms of the number of bubbles burst/missed and the number of repetitions completed. 
Anomaly can also be detected by comparing the recorded reference skeleton data with the live user skeleton data. 
For our case studies, we detect three types of anomalies:
\begin{itemize}
	\item During specific games, users may trip and \textbf{fall} in real space, potentially harming themselves. This is a serious anomaly and such occurrences should be recorded to avoid future harm, in case the object movements assisted in the fall.
	\item We check the user's \textbf{orientation} to ensure that the user is facing the station area within 90 degrees of either direction.
	\item We evaluate the hand \textbf{positioning} to ensure that the user handles the equipment properly.
\end{itemize}
Every anomaly is reported on the screen as real-time feedback to the user, allowing the user to adjust and correct their posture and movement while utilizing tools or moving objects. 
The anomalies are recorded for future SME reference and analysis. 
If any anomaly lasts for more than the anomaly wait time, e.g. $10$ seconds, we assume that the user is pre-occupied with something external, the skeleton is recorded incorrectly, or the user is having some other difficulty. 
The game is exited to remove risk of harm to the user and to save on recording storage.

\begin{figure*}[]
	\centering
	\includegraphics[width=160mm, keepaspectratio=true]{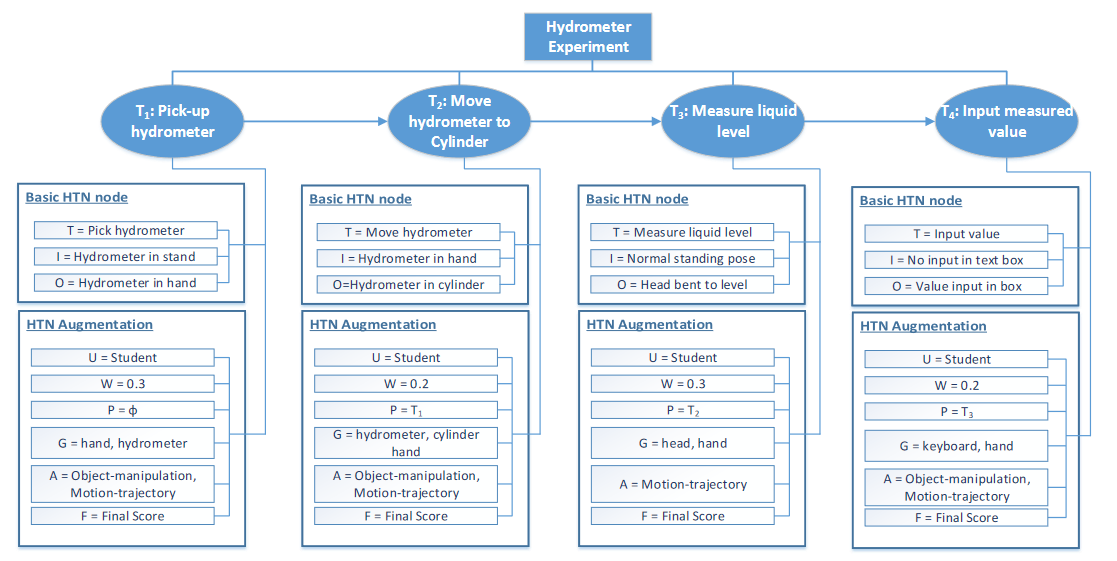}
	\caption{Assessment HTN (A-HTN) for Hydrometer Experiment}
	\label{fig:hydroHTN}
\end{figure*}


\section{CASE STUDIES} \label{caseStudies}
To evaluate the effectiveness of the A-HTN, we designed two serious game case studies: a hydrometer experiment and a collaborative chemistry experiment. 
Both experiments were developed with the Unity3D game engine using OpenGL rendering and PhysX for physics interaction.
Each performance site had one laptop with an Intel i7 3.0 GHz processor, 16 GB RAM and 4 GB GTX 1070 graphics card.
The hydrometer experiment used an Oculus Rift as the display, whereas the collaborative chemistry experiment used a 3D TV.
A Leap Motion was used to capture the hand finger joints for interaction with the virtual objects in the hydrometer experiment. 
Since the complete body of the user was captured for the collaborative chemistry experiment, we used a Microsoft Kinect V2. 
The code for generating the 3D human model was implemented in C++ with GPU optimization on Windows. 
We categorize the user study and evaluation into 2 groups depending on the type of assessment performed - task-level manipulation assessment and action-level assessment. 
Both the hydrometer experiment and the collaborative chemistry experiments included both types of assessment .
A video for task performance assessment in these serious games is available at \url{https://youtu.be/Z0z20zUTELs}.

\subsection{Hydrometer Experiment} \label{caseStudy1}
A hydrometer is an instrument that is used to measure the specific gravity (relative density) of liquids, i.e., the ratio of liquid density to water density.
We created a virtual chemistry laboratory to perform the hydrometer experiment, as shown in Figure \ref{fig:caseStudies}.
For successful completion of the experiment, the following tasks were performed:
\begin{itemize}
	\item \textit{$T_1=$ Pick-up hydrometer} from the upper part of the stem to avoid the transfer of oil from the hand to the hydrometer.
	\item \textit{$T_2=$ Move hydrometer to cylinder} by making sure it does not drop from the hand or collide with the cylinder.
	\item \textit{$T_3=$ Measure the liquid level} after the hydrometer stabilizes in the cylinder. To measure correctly, the student must bend and bring his/her head to liquid level.
	\item \textit{$T_4=$ Input measured value} in the on-screen text box slot using the keyboard.
\end{itemize}

Figure \ref{fig:hydroHTN} shows the A-HTN with the augmented node parameters for task performance assessment. 
$U=$ \textit{single user}, i.e., a student had his/her \textbf{task-level assessment}  conducted based on the $A=$ \textit{object manipulations} in the experiment as well as his/her \textbf{action-level assessment} conducted based on the $A=$ \textit{motion trajectories}.
As this experiment was performed as a part of the syllabus for a community college, $F=$ \textit{final-score} was provided as feedback.
The predecessor tasks $P_i$ were provided for each task. 
These tasks defined the sequential task performance order and are represented as arrows. 
Game objects $G=$ \textit{hydrometer, cylinder, hand, head/eye, text value input} were manipulated and compared with the SME reference performances.
\begin{itemize}
	\item For task $T_1$, the user's hand \textit{orientation} was checked to ensure that the student handled the hydrometer in the correct manner, i.e., from the upper part.
	\item For task $T_2$, an \textit{attachment} check was performed to ensure that the hydrometer was connected with the hand while it was being moved to the cylinder. A \textit{collision} check was also performed between the hydrometer and the cylinder.
	\item For task $T_3$, a \textit{position} check was performed on the student's head by comparing it with the meniscus level of the liquid in the cylinder.
	\item For task $T_4$, a \textit{text value input} check was performed by comparing the student provided liquid value with the recorded liquid value provided by the SME.
\end{itemize}

Data from these task-level manipulations were collected as evidence to address the competency of the users at understanding the density of various liquids. 
Motor skills in virtual reality games add an extra level of complexity to automatic assessment for the user as solutions are not simple click or button based actions and thus, action-level manipulations were also collected.
By following the ECD framework, we intend to measure the (1) competency of the user by having them perform (2) tasks and by analyzing the (3) evidence of the tasks. 
Specifically, we are measuring (1) competency of the user with respect to their knowledge about density and equipment handling safety while performing the hydrometer experiment.
We collect the data from (2) the task-level manipulations and the action-level trajectories and use the data as (3) evidence of the competency in both knowledge and safety. 
Utilizing the ECD framework allows us to achieve a more fine-grained understanding in what knowledge the user may be lacking while performing the experiment.

For each task, a weight $W_i$ was provided and a score $\Omega_i$ was calculated after comparison. 
Once the experiment was completed, an overall score of $\Delta = \sum_i W_i \cdot \Omega_i$ was computed as the weighted sum of all individual task scores.

\begin{figure*}[]
	\centering
	\includegraphics[width=160mm, keepaspectratio=true]{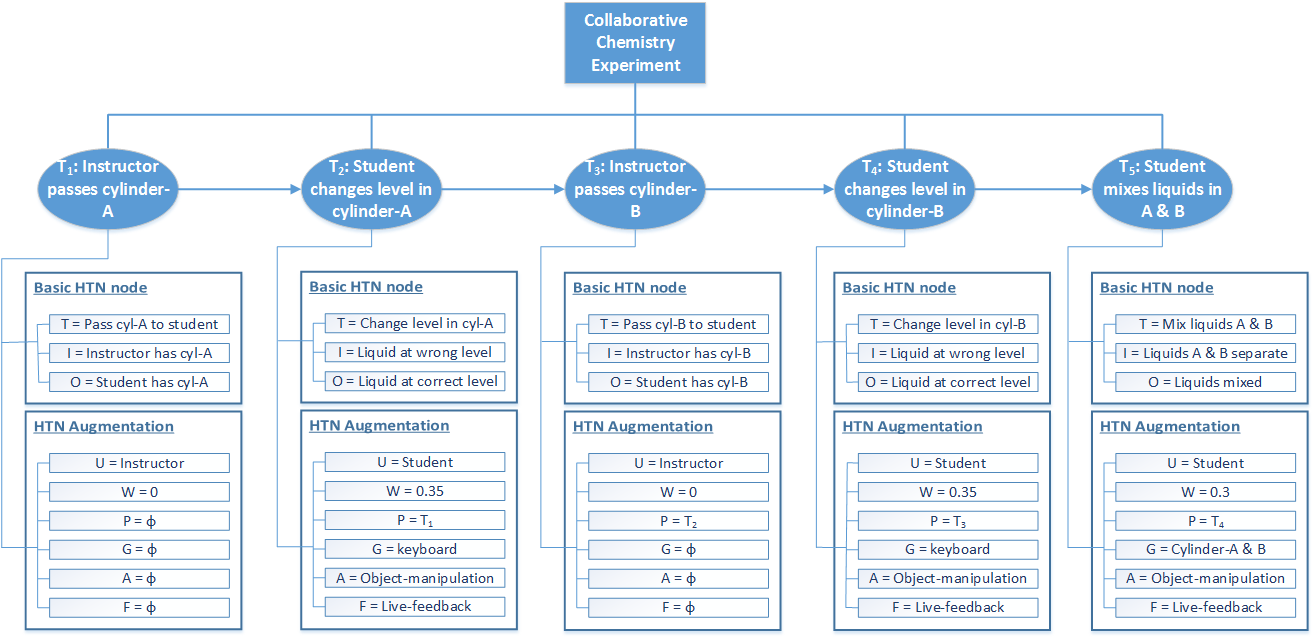}
	\caption{Assessment HTN (A-HTN) for Collaborative Experiment}
	\label{fig:collabHTN}
\end{figure*}

\subsection{Collaborative Chemistry Experiment} \label{caseStudy1}
We have shown that we can utilize A-HTN for single user 3D games. However, many environments and experiments require multiple people to be present at a common location to perform tasks together. To demonstrate this, we utilize a Collaborative Chemistry experiment, showcased in previous work \cite{kevinMMCVL} as Experiment-2.
Here, two users, an instructor and a student present in the 3D environment yet located at different physical sites, collaborate to perform a Chemistry experiment. 
In order to successfully complete the experiment, the following tasks are performed:
\begin{itemize}
	\item
	$T_1$ = \textit{Instructor passes cylinder-A} containing a liquid
	filled at a certain level to the student.
	\item
	$T_2$ = \textit{Student changes liquid level in cylinder-A} to a
	certain level as suggested by the instructor utilizing a keyboard. The
	student must verify the level of the liquid to the specified level by
	bending and measuring the cylinders in a proper manner.
	\item
	$T_3$ = \textit{Instructor passes cylinder-B} to the student once the
	liquid in cylinder-A is at the correct level.
	\item
	$T_4$ = \textit{Student changes liquid level in cylinder-B} in the
	same way as \(T_2\).
	\item
	$T_5$ = \textit{Student mixes liquids in A and B} by pouring them into
	a beaker, one after the other. Particle effect is shown for the
	chemical reaction.
\end{itemize}
A-HTN for the Collaborative Chemistry experiment is shown in Figure \ref{fig:collabHTN}. 
As shown by the arrows, the tasks for cylinder-A are performed first, after which tasks for cylinder-B are performed and lastly, their liquids are merged. 
The game objects modified in the tasks are \textit{G = cylinder-A, cylinder-B, beaker, head/eye, keyboard}. 
We utilize this experiment to demonstrate individual in a group assessment and assess performance for one user, \textit{U = student}. 
The weights of the tasks reflect the decision to assess the student by weighting $T_2$, $T_4$ and $T_5$ with $W_i > 0$.
In $T_2$, the student is required to set the liquid level in cylinder-A to the correct level before moving on. 
To accommodate this task, \textit{F = real-time} feedback is provided to simultaneously assess the student and instruct on how to perform the task correctly. 
Task-level manipulation assessment is conducted for the student's performance by comparing the \textit{A = object manipulations} with the SME reference data.

\begin{figure*}[]
	\centering
	\subfloat[$T_1$]{\includegraphics[width=55mm, keepaspectratio=true]{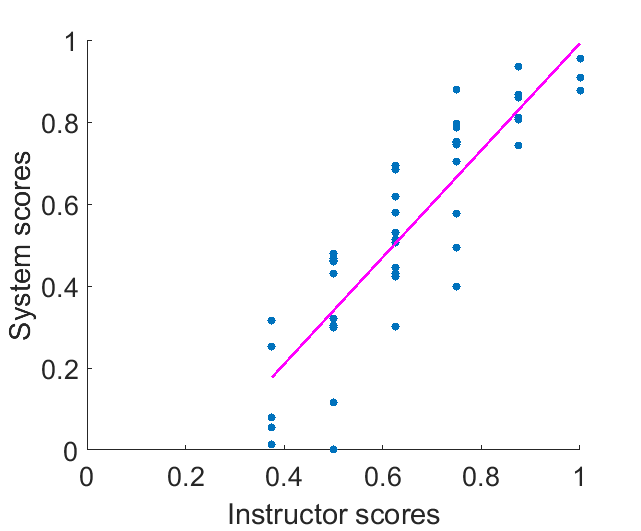}}
	\subfloat[$T_2$]{\includegraphics[width=55mm, keepaspectratio=true]{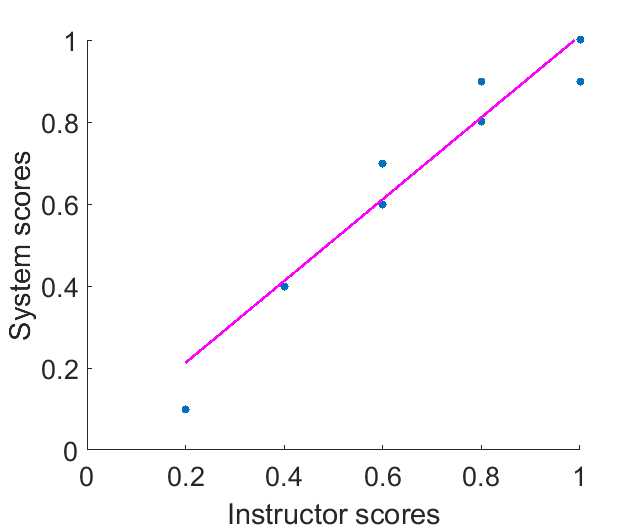}}
	\subfloat[$T_3$]{\includegraphics[width=55mm, keepaspectratio=true]{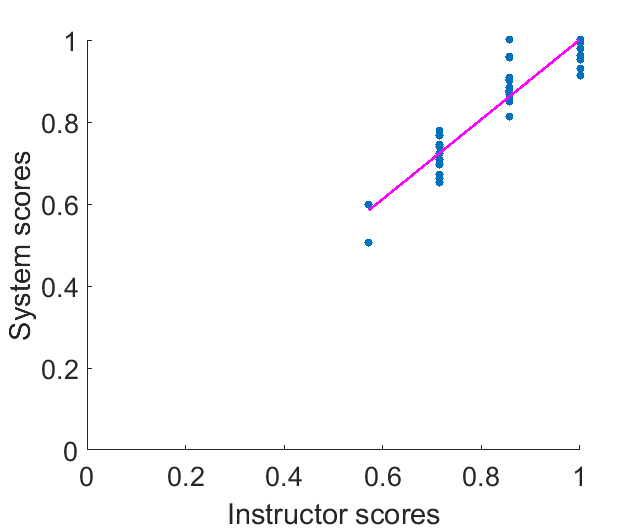}}\\
	\subfloat[$T_4$]{\includegraphics[width=55mm, keepaspectratio=true]{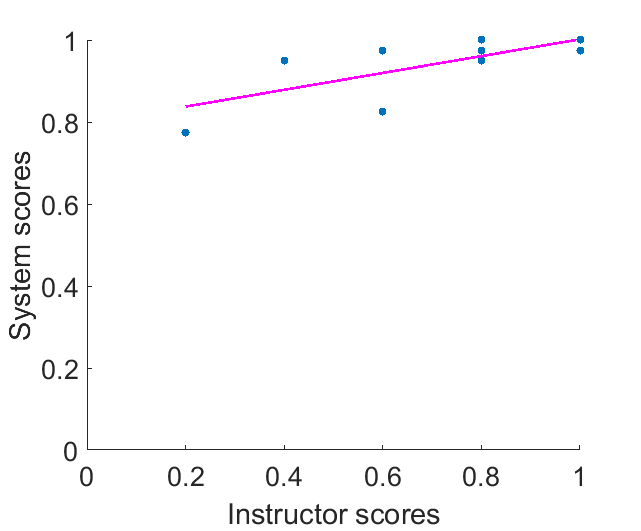}}
	\subfloat[$\Delta$]{\includegraphics[width=55mm, keepaspectratio=true]{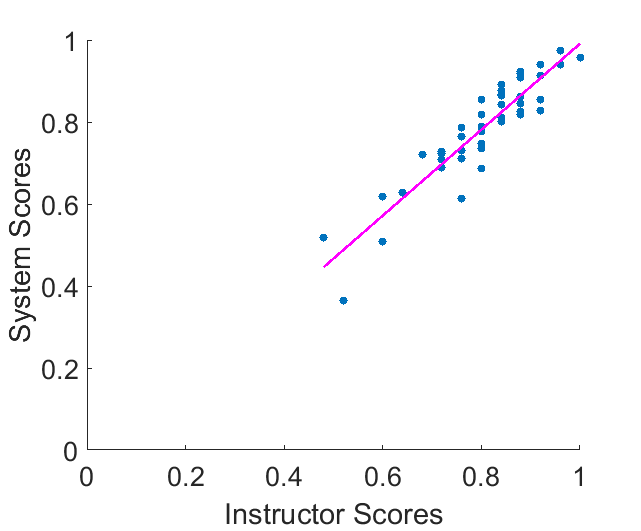}}
	\caption{Hydrometer experiment: Scatter plots for the system and instructor scores individually for (a - d) 4 primitive tasks $T_1$ - $T_4$ and (e) final assessment score $\Delta$.}
	\label{fig:corr-ind}
	\vspace{-2mm}
\end{figure*}

\section{EXPERIMENTAL RESULTS}
 \label{hydroStudy}
The \textit{hydrometer experiment} was conducted as part of a laboratory class of 45 students at a local community college under the supervision of an experienced Chemistry Professor as the SME. 
The setup required a laptop, Oculus Rift, and a Leap Motion. 
The reference data for the comparison and performance assessment were provided by the SME. 
Each student performed the experiment and had his/her task and action performance assessed, both by the instructor and the system. 
The students were first allowed some time in the sample VR room to become comfortable. 
For each student, the liquid specific gravity was selected randomly, and the color was changed accordingly. 

\begin{table}[b]
	\centering
	\caption{Hydrometer experiment: Correlation between system generated and instructor scores} \label{tab:correlation}
	\begin{tabular}[]{|c|c|c|c|}
		\cline{1-4}
		\textbf{Task} & \textbf{Pearson} & \textbf{Spearman} & \textbf{Kendall} \\ \hline
		$T_1$ & 87.54 & 88.12 & 75.28 \\ \hline
		$T_2$ & 95.45 & 89.11 & 85.41 \\ \hline
		$T_3$ & 93.78 & 91.77 & 81.95 \\ \hline
		$T_4$ & 82.56 & 85.45 & 82.00 \\ \hline
		\textbf{$\Delta$} & 91.83 & 90.69 & 77.64 \\ \hline
	\end{tabular}
\end{table}

Task-level object manipulation and action level motion trajectory assessments were performed to evaluate the students’ competency with understanding the hydrometer experiment and the safety requirements.
As the feedback type $F$ is \textit{final score}, five different scores were computed both by the instructor and by the system, one for each of the four tasks and an overall final score.
Figure \ref{fig:corr-ind}  shows five different scatter plots depicting the normalized correlation between the system generated scores and the instructor scores as a line fit.
Without assuming linearity or monotonicity between the final assessment scores provided by the system and the instructor, we report three correlation coefficients - Pearson, Spearman and Kendall, as shown in Table \ref{tab:correlation}.
All values were calculated using the MATLAB functions provided with the statistics and machine learning toolkit. 
It is clear from the values that the system generated scores correlate well with the instructor scores.
Regarding $T_1$, after review from the SME, it appears that one of the causes for variance in scores may have been caused by differences in how the user retrieved the hydrometer from the stand. 
Our system may have graded on the exact matching the joint structure, leaving little room for variation and thus, judging more harshly. 
However, the instructor felt satisfied with the overall assessment provided by the system and agreed that grading work can be significantly reduced in future.

The \textit{collaborative chemistry experiment} was also conducted as part of the laboratory class at the same community college as the \textit{hydrometer experiment}.
The experimental setup was similar to the hydrometer experiment, differing in that two sites were required instead of one. 
The chemistry instructor was present at one site and each one of the 31 students participated at the other site
Object manipulation assessment was performed for each student and real-time feedback was provided on-screen.
The feedback was visible to both instructor as well as students. 
The instructor could offer additional help using voice and visual communication if necessary. 
However, the instructor found the proposed system to be very intuitive and accurate in providing the correct feedback and rarely found the need to intervene.

\section{DISCUSSION} \label{discussion}
To the best of our knowledge, very few models and systems exist for task performance assessment in 3D serious games. 
The majority of frameworks, such as \cite{sietteAssess} and \cite{moocAssess}, focus only on the assessment of online 1D (text) or 2D (image) content.
Some detailed and effective frameworks are tailored to work for specific case studies \cite{searchAlgo2dEval}.
Experiments such as the hydrometer experiment need students to work the depth in order to safely place the fragile hydrometer in the cylinder without hitting the walls.
The collaborative chemistry experiment shows the need to assess students individually even when part of a group. 
The need for these 3D virtual environments is increasing, which eventually increases the need for automated task performance assessment.

Assessment in 3D serious games have only been considered in a handful of works. 
\cite{dbnAssess} showcased an assessment system for 3D serious games that is most related to our work. 
The system assesses task performance in a fire simulation study using a Dynamic Bayesian Network (DBN).
The work showcased are significant steps towards automated assessment in 3D games, yet they are tailored for specific use case scenarios.
While other generalization strategies have been used for various datatsets, \cite{robustSA}, we extend these models to generalize use cases for not only 3D games, but also virtual reality environments.
By utilizing the A-HTN model, we can incorporate the task and action performance of many, if not all types of 3D serious games, both virtual and non-virtual. 
These serious games include but are not limited to educational laboratories and exergames.
Due to the inclusivity of both task-level and action-level assessments, there are not any other systems that can be readily or directly compared with our proposed A-HTN. 

Another significance of our work is that we measure not only the competency of the user’s knowledge regarding the experiment, but their competency at following safety precautions as well.
For example, imagine a chemistry setting where the researcher uses a chemical such as 2-Mercaptoethanol. 
It is not only important to capture and assess how they use the chemical, but the relation of the chemical to the face. 
2-Mercaptoethanol can harm the mucous membranes and cause larynx spasms, pneumonitis, and pulmonary edema when inhaled. 
In order to capture this in a virtual setting, the user is also assessed on their motion-trajectory via action-level assessment with regards to the distance between the hydrometer and the face.

\subsection{Multi-user or Group Assessment}
Several 3D serious games allow multiple users to be present at different sites and still perform the task collaboratively. 
One such example is our Collaborative Chemistry experiment, where 2 students are present at different sites and are collaborating to perform a Chemistry experiment.
Depending on the task, the SME can select who is supposed to be assessed - $Student_1$, $Student_2$ or the entire group. 
Our proposed A-HTN can easily consider any of the 3 cases.
For each task $T_i$ , the value of the variable $U_i$ can be changed to the user/group whose task performance is being assessed.
We demonstrate the ability to evaluate individuals in a group in our experiment. 
However, by changing the weights for the instructor, we show that our proposed A-HTN captures the group score as well.

\subsection{Limitations and Future Work}
In our work, performance can be evaluated using - task-level assessment (object manipulation) and action-level assessment (motion trajectory). 
However, there are other games where a creative component is involved \cite{schuteCreative}.
For example, consider a physics environment where the user must transport an object over an obstacle. 
As our system utilizes a SME for both the tasklevel assessment, i.e., the object starting from position A to position B, and the action-level assessment, i.e., how the character transfers the object, creativity or deviation from the path of the SME would likely lead to lower results for the user. 
In future work, the A-HTN could have another assessment type, i.e., creativity measurement.
However, in our environment, safety comes first.
In order to ensure safety is taught and assessed, creativity is sacrificed.

At this point in time, the proposed A-HTN and the authoring mechanism do not attempt to incorporate creative components that may be involved with other environments.
We will take that up as possible future work. 
Similarly, we have not considered any non-serious (or “fun” only) 3D games, which may or may not be modeled by the A-HTN.

\section{CONCLUSION}\label{concl}
In this paper, we proposed Assessment HTN (A-HTN) for assessing task performance in 3D serious games. 
Previous research, including the HTN model, was mainly focused on modeling the tasks. 
However, using the augmentations added to the HTN, we can also incorporate the task performance assessment logic in any serious game. 
Using the A-HTN, we showcased two different assessment strategies - (1) Task-level Assessment by comparing object manipulations and (2) Action-level Assessment by comparing motion trajectories. 
The A-HTN can be used to perform task performance assessment for individual users, a group of users, or individual users in the group. 
The proposed A-HTN was evaluated using two serious game case studies, a hydrometer experiment and a collaborative chemistry experiment. 
A video for the serious game case studies is available at  \url{https://youtu.be/Z0z20zUTELs}.
The experimental results from the case study showcased the usability and effectiveness of the A-HTN for incorporating the assessment logic and performing automated task performance assessment in 3D Virtual Reality serious games.

\section{ACKNOWLEDGMENTS}
This material is based upon work supported by the National Science Foundation under Grant No. 2153249. Any opinions, findings, and conclusions or recommendations expressed in this material are those of the author(s) and do not necessarily reflect the views of the National Science Foundation.

\bibliographystyle{abbrv-doi}
\bibliography{task-qual}


\end{document}